\documentclass[superscriptaddress,aps,prd,preprintnumbers,nofootinbib,preprint,12pt]{revtex4}
\usepackage[utf8]{inputenc}
\usepackage[citecolor=blue,urlcolor=blue,unicode=true,pdfusetitle,
bookmarks=true,bookmarksnumbered=true,bookmarksopen=true,bookmarksopenlevel=3,
breaklinks=false,pdfborder={0 0 1},backref=false,colorlinks=true]{hyperref}
\usepackage{bm}
\usepackage{latexsym}
\usepackage{dcolumn,xcolor}
\usepackage[normalem]{ulem}
\usepackage{amsmath,amsfonts,amssymb}
\usepackage{graphicx,epsfig}
\usepackage{soul}
\usepackage{ulem}
\usepackage{amsthm}
\usepackage{color,float} 

\usepackage{amssymb}
\usepackage{braket}
\usepackage{physics}

\usepackage{tikz}

\begin{document}
	\title{ The end of spacetime }
 
\author{Mir Faizal}
\email{mirfaizalmir@googlemail.com}
\affiliation{\scriptsize{Irving K. Barber School of Arts and Sciences,
University of British Columbia - Okanagan, Kelowna, British Columbia V1V 1V7, Canada}}
\affiliation{\scriptsize{Canadian Quantum Research Center 204-3002  32 Ave Vernon, BC V1T 2L7 Canada}}

	\date{\today}
	\begin{abstract}
 We will highlight that despite there being various approaches to quantum gravity, there are  universal 
 approach-independent features of quantum gravity.  The geometry of spacetime  becomes an  emergent structure, which emerges from  some purely quantum gravitational degrees of freedom. We   argue that these quantum gravitational degrees of freedom can be best understood using quantum information theory.  Various approaches to quantum gravity seem to suggest that quantum gravity could be a third quantized theory, and such a theory would not  be defined in spacetime, but rather in an abstract configuration space of fields. This supports the view that spacetime geometry is not fundamental, thus effectively ending the spacetime description of nature. 
 	\\
		{}
		\\
		{}\\

	\end{abstract}
	\maketitle
\newpage
In Newtonian mechanics,   space and time occur as  distinct physical concepts, and events occur at a place in space and at a point in time. This concept does not change with respect to who is observing those events. In this sense, Newtonian mechanics  makes  space and time an  absolute objective observer-independent  concept \cite{n1}.
Special relativity challenges this notion of space and time and merges them into a single physical quantity, which is usually called spacetime \cite{rela}. This concept of spacetime still remains  an  absolute objective 
observer-independent  concept, but both space and time cease to exist as independent concepts in special relativity. If taken individually, they explicitly depend on the relative motion of the person making the measurement. Despite space and time losing independent meaning, the geometrical structure of spacetime is flat and fixed in special relativity. General relativity makes spacetime dynamic, and according to general relativity, spacetime curves in presence of matter \cite{gene}. In fact, it is this very curvature of spacetime that causes gravity to act on objects. Among other things, proof of this concept occurs by correctly predicting the bending of light by stars \cite{light}. According to Newtonian gravity, only massive particles can be affected by gravity. As light classically is a  massless wave (neglecting the quantum theory of light), it should not have been affected by the gravitational force.  Now according to special relativity, light has to travel in a straight line, so it should have continued to do so even in the neighborhood of stars. However, its bending near stars, demonstrates that gravity is actually the curvature of spacetime, and due to the curvature of spacetime, a straight line is actually a curve.

 In both special and general relativity, even though space and time are subjective notions, spacetime is an objective observer-independent  concept. However, the very concept of objective observer independence has to be totally discarded in quantum mechanics \cite{quan}. Quantum mechanics is not a mere probabilistic theory, like statistical mechanics. In quantum mechanics, the very act of measuring any event changes the system. As an observer performs these  measurements, it becomes impossible to define reality in an objective 
 observer-independent way. However, apart from the measurement problem \cite{meas}, which only becomes important in purely quantum phenomena like entanglement \cite{enta} (a quantum property by which two quantum particles remain connected even when separated by vast distances, and causes purely quantum effects described by the Bell's theorem), quantum mechanics for all practical purposes can be reduced to a statistical theory. This is because, even though we cannot predict the outcomes of individual experiments, we can assign probabilities to them, and these probabilities match with the observations for a larger number of such measurements. This deep relation between quantum averages and statistical averages is the reason quantum mechanical tools like path integrals can be used as calculation devices in statistical mechanics \cite{path}. But despite the mathematical similarity between classical statistical mechanics and quantum mechanics, these theories are conceptually very different from each other. 

It has been possible to reconcile special relativity with quantum mechanics, and the theory which does this is called  quantum field theory \cite{qft}. Conceptually, it is known that if we naively try to write a relativistic generalization of the Schrödinger equation, we get inconsistencies. These inconsistencies known to occur in the  Klein-Gordon equation could be resolved in the Dirac equation. However, if we treat even the  Dirac equation as a quantum mechanical equation, we cannot account for the   dynamical creation and annihilation of particles, which occurs in quantum field theory. The deep relation between the quantum systems and statistical mechanics is used to first express the relativistic equation like the Klein-Gordon equation as a 'classical' equation of a classical field, and then it is quantized again \cite{seco}. The quantum mechanical observer dependence is absorbed in this second quantization of the field, and the original  Klein-Gordon equation is treated as a classical deterministic equation. The advantage of doing this is that now it becomes a quantum theory of a field, where its perturbative modes act as particles, and hence the particles can be created and annihilated by any non-linear terms involving those fields. Thus,  elementary particles (such as electrons, photons, etc) of nature stop being fundamental (as irreducible building blocks) as they can be produced due to the collision of other particles.  It may be noted that this is only possible in second quantized theories, with non-linear terms. In first quantized theories, these particles are truly fundamental, and cannot transform into different particles. 
Furthermore, it is possible to even reconcile quantum mechanics, with general relativity, if spacetime is not quantized. This is done using quantum field theory in curved spacetime \cite{qftc}. In this theory,  a stationary observer does not observe any particles, whereas an accelerating observer observes particles at a certain temperature \cite{un, un12}. This effect called the Unruh effect has even been experimentally observed \cite{un14}. Thus,  the very notion of particles also becomes an observer-dependent concept and depends on the acceleration of an observer. 

 
However, it has not been possible to fully unify quantum mechanics with general relativity, in a complete theory, where even spacetime will be quantized. Even though we do not have a fully consistent quantum theory of gravity, we have several approaches to quantum gravity. The interesting observation is that these approaches seem to make certain universal predictions about the nature of quantum gravity. To investigate these universal predictions, we observe that  we need more energy to  probe spacetime at smaller lengths scales. We need Planck energy to probe a Planckian region. However, if we put Planck energy in the Planckian region, we will form a mini black hole, which will in turn prevent such measurement from taking place \cite{bl}. Now any theory of quantum gravity has to be consistent with classical black hole physics. Thus, it seems that in any quantum theory of gravity, spacetime should have a natural minimal length \cite{le} and a minimal time \cite{ti}. This would also solve another problem in quantum field theories. In quantum field theories, we get divergences or infinite answers. To get finite answers, we need to introduce a cutoff by hand, which is a scale beyond which we do not probe our system. However, if spacetime does have a minimal length and a minimal time, it will naturally act as a cutoff for any quantum field theory described on such a spacetime \cite{cut}.  
 
Perturbative string theory does have such a minimal length, as it is not possible to probe spacetime below string length scale \cite{pert}. This is because in perturbative string theory, string length is the smallest probe available, and we cannot probe spacetime below the length of the smallest probe. Then taking the foundations of quantum mechanics seriously, any physical object that cannot be probed does not physically exist, so string theory suggests that spacetime does not exist below the string length scale. Even though we have point-like objects known as D0-branes in non-perturbative string theory, this conclusion still holds because of dualities in string theory. Due to a duality,  called T-duality, between string theory at larger and small scales, it has been demonstrated that the spectrum of string theory above string length scale is similar to the spectrum of string theory below string length scale \cite{tdual}. Hence, new information is not obtained by probing string theory below the string length scale, and this length scale acts as a minimal length. Similarly, a minimal length occurs in other discrete models of spacetime, like the   causal sets \cite{c1},  quantum graphity \cite{c2}, and causal dynamical triangulation \cite{c4},  where it is not possible to probe the geometry below the length scale at which discreteness occurs. Thus, the occurrence of a minimal length, or more generally a minimal geometrical quantity as a natural cutoff of spacetime seems to be an approach to independent observation in quantum gravity. Such a minimal length also occurs in noncommutative \cite{nc10} and nonlocal \cite{nl10} quantum field theories. In loop quantum cosmology a background independent quantized called the polymer quantization has to be employed \cite{c5, c51}. In this quantization, there is a natural minimal length called the polymer length. Apart from that, a minimal area and a minimal volume occur naturally in  loop quantum gravity \cite{c6} and spin foam \cite{c10}. Thus, we still have a minimal value for some geometric quantity, which acts as a natural cutoff for  geometry. Such a cutoff seems to be a generic feature of any theory of quantum gravity, even though the details of its origin are approach dependent.  


Such a  cutoff for  geometry resolves some intrinsic problems in general relativity.   General relativity predicts its own breakdown due  to the occurrence of singularities, where spacetime description of reality does not hold. Furthermore, the Penrose-Hawking singularity theorems demonstrate that these singularities are an intrinsic property of general relativity \cite{ph1, ph2}. However,  quantum gravitational effects should modify these theories by including a cutoff on geometry and  prevent the occurrence of these singularities \cite{li1, li2}. In string theory, a minimal length occurs due to T-duality \cite{tdual}, and T-duality also prevents the formation of  singularities \cite{tdual1}. Singularities are also removed in loop quantum cosmology due to  the cutoff on the geometry \cite{lqgs}.  As the geometry in any theory of quantum gravity has a natural cutoff, it can be argued that the absence of  singularity  will be a general feature of all approaches to quantum gravity. This is because the  Penrose-Hawking singularity theorems  have been directly related to the Bekenstein-Hawking entropy \cite{hpbe}, so a bound on such an entropy from the cutoff would naturally prevent the occurrence of singularities. 
As the   Bekenstein-Hawking entropy is directly related to geometry in the Jacobson formalism \cite{jc}, a modification to the Bekenstein-Hawking entropy  will directly modify the geometry of spacetime.   In fact, it  has been explicitly demonstrated  that the bound on the  Bekenstein-Hawking entropy  from the minimal length in quantum gravity will  prevent the formation of  singularities in spacetime \cite{gup, gup2}.  The singularities are prevented as a minimum value for a geometric quantity would imply that spacetime geometry is an emergent structure, which emerges from quantum gravitational degrees of freedom, and breaks down below such a minimal value. Singularities occur in general relativity when we apply it to describe scenarios, where the spacetime description of nature does not hold. It is also interesting to note that this bound on the geometry follows most directly from a bound on the quantum information  \cite{gup, gup2}, which seems to indicate that spacetime geometry emerges from quantum information. 

The emergence of spacetime geometry from purely non-geometric quantum states can be most explicitly seen using the holographic principle. The   holographic principle relates a theory in a volume of a region to the theory on its boundary  \cite{holog}.
The holographic principle has been thoroughly applied to study various aspects of string theory. Here, the holographic principle relates a gravitational theory to the quantum field theory on its flat boundary \cite{ads}.  Thus, it is studied as a duality between string theory, or more commonly its supergravity approximation on anti-de Sitter spacetime (spacetime with a constant negative scalar curvature) and a boundary  conformal field theory  (a theory where angles are  preserved   but not lengths).    The emergence of spacetime geometry from non-geometric quantum degrees of freedom follows directly from this form  of the  holographic principle  \cite{ep1, bits, b12, b14}.   This has been done by relating the geometric structures  to the  abstract notion of   information, which has been made concrete in quantum  information theory. As discussed before, even though most quantum mechanical phenomena can be mapped onto statistical mechanical systems, there are purely quantum effects like entanglement  for which this cannot be done. Interestingly, it is the entanglement of the quantum states in the boundary theory which gives rise to the geometric structure. Removing entanglement in the boundary theory amounts to destroying the geometric structures for  dual theory. However, entanglement has been demonstrated to be an observer-dependent phenomenon \cite{obser, obser1, obser2, obser4}. This also makes the geometric structures dual to such entanglement   also observer-dependent.  

In loop quantum gravity \cite{c6} and spin foam \cite{c10}, geometry at macroscopic scales  emerges from   loops woven into a   network at the microscopic scale. Here, again the geometry of spacetime emerges only  as an approximation at a larger scale compared to the Planck scale, much like the geometry of a table emerges from atomic physics at a larger scale compared to the atomic scale.  This also occurs in other models of quantum gravity, with discrete degrees of freedom, such  as causal sets \cite{c1},  quantum graphity \cite{c2}, and causal dynamical triangulation \cite{c4}. Thus, the idea that geometry  emerges from non-geometric states seems to be a universal prediction of various approaches to quantum gravity.   Furthermore, even though these other approaches to quantum gravity are fundamentally different from string theory, even in those approaches, geometry seems to emerge from quantum  information theory   \cite{Makela:2019vgf, bombelli, konopka, Skenderis:2006ah}.


The idea of emergence can be made dynamical in third quantized theories \cite{t1}. Just like in the second quantized theory, it is possible to dynamically create and annihilate particles, it is possible to define a theory, where it is dynamically possible to create and annihilate fields, and even spacetime \cite{t2}. This theory naturally is not defined in spacetime, but in an abstract configuration space of fields, including the metric, which accounts for the geometry. Then the wave function of  quantum field theory, which is defined for fields,  is again treated like a classical  field, and a second quantized theory is defined as a classical theory of an ensemble  of fields. This theory is   quantized  again, and the measurement problem is absorbed in the third quantization of fields. Now it is possible to dynamically create and annihilate those fields, including metric, and hence it is possible to dynamically create and annihilate geometries.  Various, approaches to quantum gravity, if taken to their logical conclusion tend to end in some sort of a third quantized theory. In loop quantum gravity along with some other models of quantum gravity, third quantization has appeared in the form of group field theory, where the group theoretical structure of certain groups  is the  fundamental variable from which geometry emerges \cite{lt}. In string theory, third quantization has appeared as string field theory \cite{st02}, even though for historical reasons string theory is called a first quantized theory (despite being a second quantized conformal field theory), and string field theory is called a second quantized theory (despite being a third quantized theory, with dynamical creation and annihilation of fields).

Apart from quantum gravity, the calculations of particle scattering amplitudes also suggest that spacetime geometry is an emergent structure, and physical processes such as scattering amplitudes  can be better understood without reference to spacetime \cite{ar}. 
It seems natural even from a purely cosmological perspective that spacetime should be emergent. As the universe, which is represented by geometry, was formed at the big bang, and so geometry could not be fundamental but emergent \cite{cos1, cos2}. So,   in all  approaches to quantum gravity, spacetime geometry emerges from the purely  quantum gravitational degrees of freedom, and it is the information contained in those degrees of freedom that causes the geometrical structures to emerge.    This view that spacetime emerges from information  goes back to John Wheeler and is usually stated as 'it from bit not bit from it' \cite{itfrombit}. Here, 'it' represents geometry and structures such as matter defined on that geometry, and    'bit' represents the abstract notion of   information, from which that geometry emerges. The quantum gravitational variant of this initial proposal can be stated as  'it from qubit and not qubit from it' with the information being replaced by quantum information representing quantum gravitational degrees of freedom. This quantum informational need not be defined in spacetime, and in third quantized theories, it is not defined in spacetime. Furthermore, various theories of quantum gravity, such as string theory, and loop quantum gravity, along with some other approaches to quantum gravity are better understood using a third quantized formalism. Here, we would like to also remark that the spacetime in string theory is continuous despite there being a minimal length, as the minimal length just sets a limit to the ability to probe the system. However, this is equivalent to a discrete spacetime  from an information theoretical perspective, as it has been argued that like the information in information theory, spacetime   can   be simultaneously described as a discrete and continuous structure \cite{cutoff2}. Thus, from an information theoretical view, these different approaches to quantum gravity might be different representations of the same or some very similar theory. Thus, the universal feature of fundamentally different approaches to quantum gravity is the emergence of spacetime.  The  journey, which started with Newtonian mechanics, with the definition of space and time as  an absolute objective  observer-independent concept, seems to have reached a point, where the very concept of spacetime has ended.

\end{document}